\documentclass[12pt]{article}
\usepackage{epsfig}
\usepackage{amsmath}
\usepackage{hhline}
\usepackage{amssymb}
\usepackage{times}
\usepackage{cite}

\newlength{\dinwidth}
\newlength{\dinmargin}
\begin{document}  

\begin{titlepage}

\vspace*{1.5cm}

\begin{center}
\begin{LARGE}
\mbox{\bf Investigation of Quark--Antiquark Interaction }

\vspace*{0.25cm}
{\bf\boldmath Properties using Leading Particle Measurements }

\vspace*{0.4cm}
{\bf\boldmath in $e^+e^-$ Annihilation }
\end{LARGE}

\vspace{1.7cm}

{\large Martin Erdmann} \\

\vspace{1.2cm}

\noindent
Institut f\"ur Experimentelle Kernphysik,
Universit\"at Karlsruhe, \\ Wolfgang-Gaede-Str. 1, 
D-76131 Karlsruhe \\
Martin.Erdmann@cern.ch

\end{center}

\vspace{1.2cm}

\begin{abstract}
Measurements of heavy quark production in electron--positron collisions
are used to analyse the strong interactions between quarks and 
anti-quarks.
A scaling behaviour is observed in distributions of the rapidity change of 
D$^*$, B$^*$, and B mesons. From 
these distributions information is obtained on the hadron formation time, 
effective quark masses, and the potential between quark--antiquark pairs.
Predictions for fragmentation functions are presented.
\end{abstract}

\vspace{2.5cm} 

\noindent
PACS: 13.65.+i \\ \\
Key words: \\
{\em fragmentation, formation time, quark masses, 
strong interaction potential} 
\end{titlepage}

\section{Motivation}

\noindent
Measurements of fragmentation processes imply that quarks are subjected
to large energy losses from strong interactions before they form hadrons.
The production of heavy
quark--antiquark pairs at electron--positron colliders 
provides a suitable environment 
in which to analyse these strong interactions, namely,
the initial state is colour neutral, and the initial kinematics is 
known from the beam energy.
The final state momenta of the quarks are carried away by the measured
D and B mesons, except for corrections due to the 
second quark which is required for hadron formation and 
higher meson resonances.

The perspective followed in this analysis is that the primary quark 
loses energy during the time it remains in the strong field originating from 
the primary quark and anti-quark travelling away from each other.
The questions addressed are: 
how long did the quark remain in the field~? 
are there variations in the field strength~?
what is the influence of the quark mass~? 
The answers will be used to predict fragmentation functions for 
particles which contain the primary quark and which have
not decayed.

\section{Rapidity Change \label{sec:rap}}

\noindent
Instead of the fractional energy $x=E_{hadron}/E_{beam}$, commonly
used in the studies of fragmentation data, the rapidity change
$\Delta y = y_i - y_f$ in the direction of motion is used in this analysis.
We take advantage of the fact that $\Delta y$ is the same in the laboratory
and in the rest frame of the quark.
The initial rapidity $y_i$ of the quark with mass $m$ is given by the
$e^+e^-$ center of mass energy $\sqrt{s}$ via $y_i=\ln{(\sqrt{s}/m)}$.
The quark final state rapidity is $y_f$ and corresponds to that
of the primary hadron.
Throughout this analysis
\begin{equation}
\Delta y = - \ln{x}
\label{eq:dylnx}
\end{equation}
will be used where $\Delta y$ is, for the data considered in this analysis,
a good approximation to the quark rapidity change as defined above.

\begin{figure}[htt]
\setlength{\unitlength}{1cm}
\begin{picture}(15.0,10.0)
\put(7.5,9.) {\large a)}
\put(10.7,9.){\large b)}
\put(10.7,5.3){\large c)}
\put(-0.5,-0.5)
{\epsfig{file=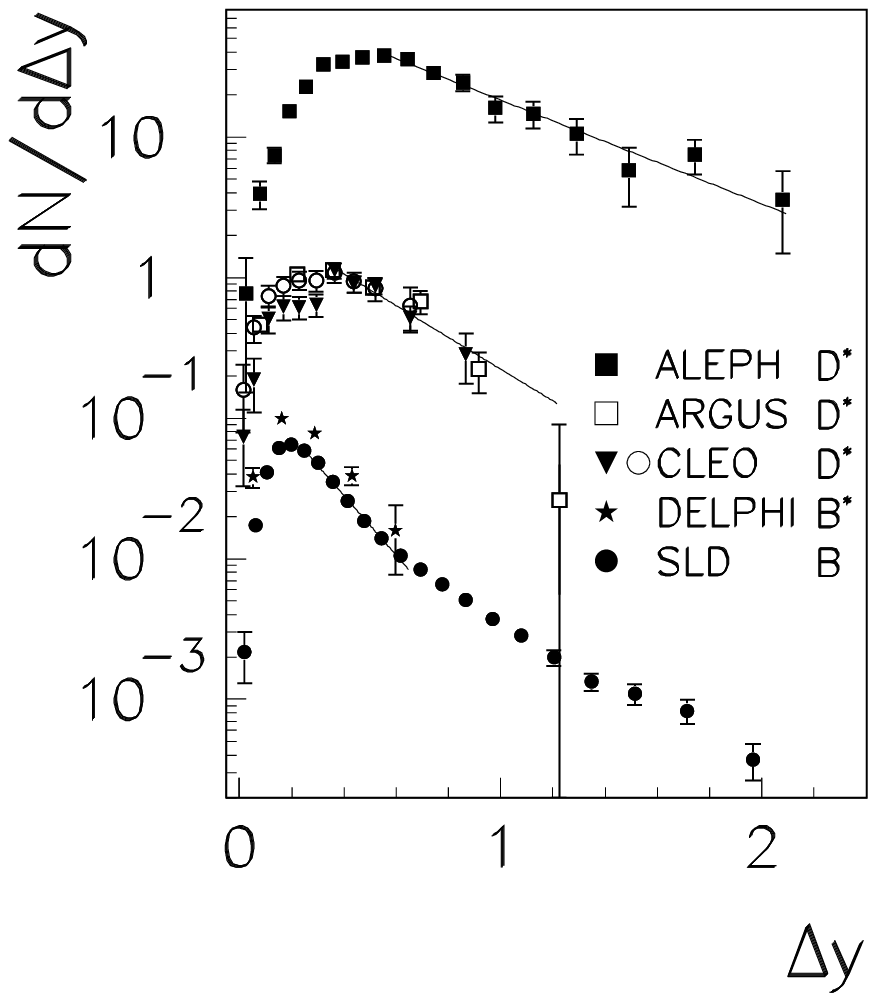,width=10.0cm}}
\put(8.0,-0.3)
{\epsfig{file=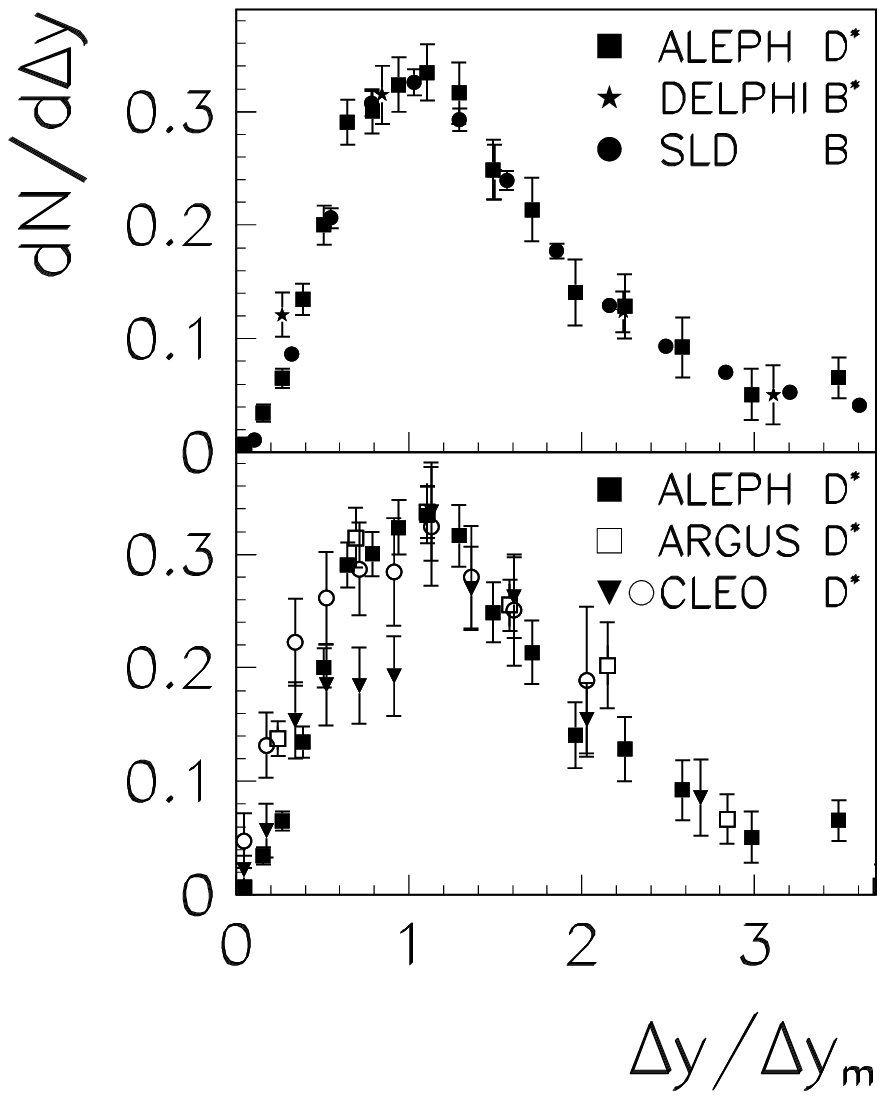,width=9.0cm}}
\end{picture}
\caption{a) Shapes of distributions of the rapidity change $\Delta y$ 
are shown for D$^*$, B$^*$, and B mesons.
The curves represent exponential fits. 
b,c)~The same spectra of the rapidity change are shown, 
normalized to the rapidity change $\Delta y_m$ where the distributions 
are maximal.}
\label{fig:dy} 
\end{figure}

In Fig.~\ref{fig:dy}a, the shapes of distributions 
of the rapidity change $\Delta y$
are shown for D$^*$ meson production in non-resonant 
$e^+e^-$ annihilation at $\sqrt{s}=10.6\;$ GeV
\cite{cleo,argus},
and D$^*$ production from $c\bar{c}$ decays of the Z boson \cite{aleph}.
Also shown is B \cite{sld} and B$^*$ \cite{delphi} meson production from 
Z decays to $b\bar{b}$.
All data exhibit a maximum at small rapidity losses $\Delta y_m$.
Above the maximum, the measurements can be described by exponentially
falling dependences.
In the case of the B meson measurement (full circles), 
two exponential components are visible in the data of which
the first is likely to reflect primary B mesons,
since the B$^*$ meson data (star symbols) show a similar dependence.
The measurements with the smaller slope at large $\Delta y$ 
are presumably due predominantly to B mesons from resonance decays and are not
further considered in this analysis.

The curves show fits to the data of the form
\begin{equation}
\left[\frac{1}{N_{norm}} \; \frac{dN}{d\Delta y}\right]_i
= A_i \; \exp{\left(-\frac{\Delta y}{\Delta y_\circ}\right)} \; .
\label{eq:expdy}
\end{equation}
A common exponential slope $\Delta y_\circ$ was determined for each 
quark flavour and $\sqrt{s}$ (Table~\ref{tab:fits}).
The normalizations $A_i$ of each data set $i$ were left free in these fits.
\begin{table}[hhh]
\begin{center}
\begin{tabular}{|l|l|l|l|l|l|l|}    
\hline
Experiment & Energy & Meson & $1/\Delta y_\circ$ & $1/\Delta y_m$ & $N_{q,\circ}/N_{norm}$ & $C$ \\
           & [GeV]      &          &                  &        &   &\\
\hline
ARGUS  & $10.6$ & D$^*$ & $2.6\pm 0.5$ & $3.1\pm 0.3$ & $1.7\pm 0.2$ & $0.30\pm 0.13$\\
CLEO   &        &       &              & & $81 \pm  10$ & $0.50\pm 0.13$\\
       &        &       &              & & $71 \pm  10$ & $0.32\pm 0.10 $\\
\hline
ALEPH  & M$_{\rm Z}$  & D$^*$ & $1.7\pm 0.2$ & $2.0\pm 0.1$ & $\left[2.28\pm 0.12\right]10^{-3}$
& $0.43\pm 0.07$
\\
\hline
SLD    & M$_{\rm Z}$  & B     & $4.9\pm 0.2$ & $5.2\pm 0.1$ & $0.935\pm 0.012$ & $0.49\pm 0.05$\\
\hline
\end{tabular}
\end{center}
\caption{Fit results of the exponential slope $\Delta y_\circ$, 
the rapidity change $\Delta y_m$ at the maximum of the spectrum,
the normalization $N_{q,\circ}/N_{norm}$, and
the correction $C$ for the start of the region where the strong
force is constant.}
\label{tab:fits}
\end{table}

The values of the rapidity change $\Delta y_m$, at which the spectra 
are maximal, 
were also determined using a suitable fit (Table~\ref{tab:fits}).
Both characteristics of the spectra, $\Delta y_m$ and $\Delta y_\circ$, 
give similar values which motivates tests of the similarity of the full
spectra.
In Fig.~\ref{fig:dy}b,c all data are shown by multiplying the rapidity
change with $1/\Delta y_m$.
The measurements at the Z resonance are then found to follow a universal
shape, independent of the quark flavour (Fig.~\ref{fig:dy}b).
At fixed quark flavour (Fig.~\ref{fig:dy}c),
the comparison of the measurements at different beam energies also
indicate similar shapes, except for the region of smallest
$\Delta y$ values.
Although the errors of the data at the lower beam energy are large,
they have the tendency to be above the measurements at the Z resonance.

The observation of a scaling behaviour 
in the spectra of the rapidity change is of interest
and is suggestive of a common hadronization mechanism.

\section{Proper Time}

\noindent
The distribution of the proper time $\tau$, measured in the rest frame of the
quark, is related to the rapidity distribution by 
\begin{equation}
\frac{dN}{d\tau} = \frac{dN}{d\Delta y} \; \frac{d\Delta y}{d\tau} \; .
\label{eq:acc}
\end{equation}
The differential $d\Delta y/d\tau$ is related to the acceleration 
of the quark.
In the case of constant acceleration, the rapidity 
distribution gives a direct measure of the proper time distribution.

The region of large $\Delta y$ in Fig.~\ref{fig:dy} corresponds to 
measurements at large distances between the quark--antiquark pair
where the quark has remained in the field for a relatively long time interval.
This long distance component of the strong interaction has been described by
a constant force of strength $\kappa \sim 1$~GeV/fm, 
e.g. \cite{eichten}, 
implying constant acceleration,
\begin{equation}
\frac{d\Delta y}{d\tau} \sim \frac{\kappa}{m} = const. 
\label{eq:con}
\end{equation}
The $\tau$-distribution, corresponding to (\ref{eq:expdy}),
is then exponentially distributed,
\begin{equation}
N_q = N_{q,\circ} \; \exp{\left(-\frac{\tau}{\tau_\circ}\right)} \; .
\label{eq:tau}
\end{equation}

Therefore, if the data at sufficiently 
large $\Delta y$ reflect the rapidity losses 
due to a constant force, the proper time $\tau$ is here exponentially
distributed.
This then implies that 
the probability for the appropriately coloured and flavoured light
anti-quark appearing and joining with the heavy quark to form the hadron 
is constant with respect to the clock of the travelling heavy quark.
A similar interpretation of $\Delta y$ spectra has been suggested 
for $p p \rightarrow p X$ \cite{bde} and $\mu N \rightarrow J/\Psi X$ 
\cite{de} data.

In what follows, we take the view that the mechanism behind fragmentation
is probabilistic in nature and can be described by (\ref{eq:tau})
with a universal formation time $\tau_\circ$ which is
independent of the strength of the force and the quark flavour.

\section{Effective Quark Masses}

\noindent
One implication of a probabilistic fragmentation mechanism 
with a universal formation time $\tau_\circ$
is the direct relation of the 
rapidity change spectra to the masses which are accelerated. From 
integration of (\ref{eq:con}) it follows
that 
\begin{equation}
m \sim \frac{\kappa \tau}{\Delta y} \sim
\frac{\kappa \tau_\circ}{\Delta y_\circ} \; .
\label{eq:m}
\end{equation}
The different exponential slopes $\Delta y_\circ$, determined from 
the fits shown in Fig.~\ref{fig:dy}, are then related to the different
effective masses accelerated in the field.
The energy scale is given by $\kappa \tau_\circ$ which is expected to
be of the order of $1\;$ GeV.

\begin{figure}[htt]
\setlength{\unitlength}{1cm}
\begin{picture}(15.0,12.0)
\put(2.0,0.0)
{\epsfig{file=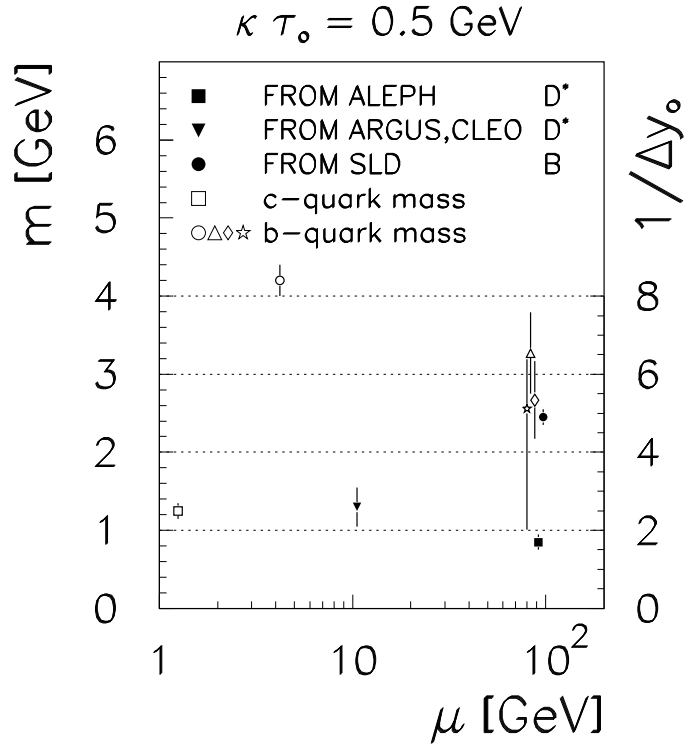,width=10.0cm}}
\end{picture}
\caption{The inverse of the exponential slope $\Delta y_\circ$ (right axis)
is shown as a function of the beam energy $\mu$ by the full symbols.
The left axis gives the corresponding masses $m$ using the energy scale
$\kappa \tau_\circ = 0.5$ GeV.
For comparison, determinations of the $c$- and $b$-quark masses are shown 
by the open symbols (see text).}
\label{fig:mass} 
\end{figure}

In Fig.~\ref{fig:mass}, the open symbols show results
on the $c$- and $b$-quark masses at the corresponding energy~$\mu$.
The open square (circle) symbol gives the value of the 
$c$ ($b$) mass of reference \cite{pdg}.
The other open symbols result from $3$-jet rate measurements in
$b\bar{b}$ events produced at the Z resonance
(\cite{delphi-bmass} diamond  symbol,
 \cite{brandenburg}  star     symbol,
 \cite{aleph-bmass}  triangle symbol).

The values of $1/\Delta y_\circ$ are shown by the full symbols in the 
same diagram.
The energy scale $\kappa \tau_\circ=0.5\;$ GeV has been chosen,
adjusting $1/\Delta y_\circ$ of the D$^*$ and B data to
the $c$- and $b$-quark mass determinations respectively.
The effective mass values of the D$^*$ data are found to be consistently below 
the B data and indicate a dependence of the charm mass on the beam energy 
scale.

In this way, the observed scaling in the spectra of the rapidity change
(Fig.~\ref{fig:dy}) can naturally be explained by different quark masses being
decelerated in the strong field.

\section{Quark--Antiquark Interactions}

\noindent
A second implication of a probabilistic fragmentation mechanism 
with a universal formation time $\tau_\circ$ is that variations in the 
strength of the force acting on the quarks can be determined as a 
function of the proper time.
Taking the factor $dN/d\tau$ in equation (\ref{eq:acc}) from 
equation (\ref{eq:tau}), 
integration of the measured rapidity spectra gives a relation between the 
rapidity change $\Delta y$ and the proper time $\tau$,
\begin{equation}
\tau(\Delta y) = - \tau_\circ \;\; \ln{\left( 1 - \frac{N_{norm}}{N_{q,\circ}}
\int_0^{\Delta y} \left[ \frac{1}{N_{norm}} \; \frac{dN}{d\Delta y} \right]
\; d\Delta y \;\right)} \; .
\label{eq:etau}
\end{equation}

Since the experiments chose different procedures to normalize their
data, the ratio $N_{norm}/N_{q,\circ}$ is determined from the 
data at sufficiently large $\Delta y$, where (\ref{eq:m}) implies
a linear relation between $\Delta y$ and $\tau$,
\begin{equation}
\frac{\tau}{\tau_\circ} = \frac{\Delta y}{\Delta y_\circ} - C \;.
\label{eq:taudy}
\end{equation}
Here $C$ accounts for the start of the region where the force is constant.
Replacing $\tau/\tau_\circ$ in (\ref{eq:etau}) by 
(\ref{eq:taudy}) allows $N_{norm}/N_{q,\circ}$ and $C$ to be determined
(Table~\ref{tab:fits}):
\begin{equation}
\int_0^{\Delta y} 
\left[ \frac{1}{N_{norm}} \frac{dN}{d\Delta y} \right] \; d\Delta y = 
\frac{N_{q,\circ}}{N_{norm}}
\left( 1 - \exp{\left(- \frac{\Delta y}{\Delta y_\circ} + C
\right)}  \;  \right)  \; .
\end{equation}

\begin{figure}[htt]
\setlength{\unitlength}{1cm}
\begin{picture}(15.0,10.0)
\put(2.4,9.) {\large a)}
\put(10.1,9.){\large b)}
\put(-0.6,-0.5)
{\epsfig{file=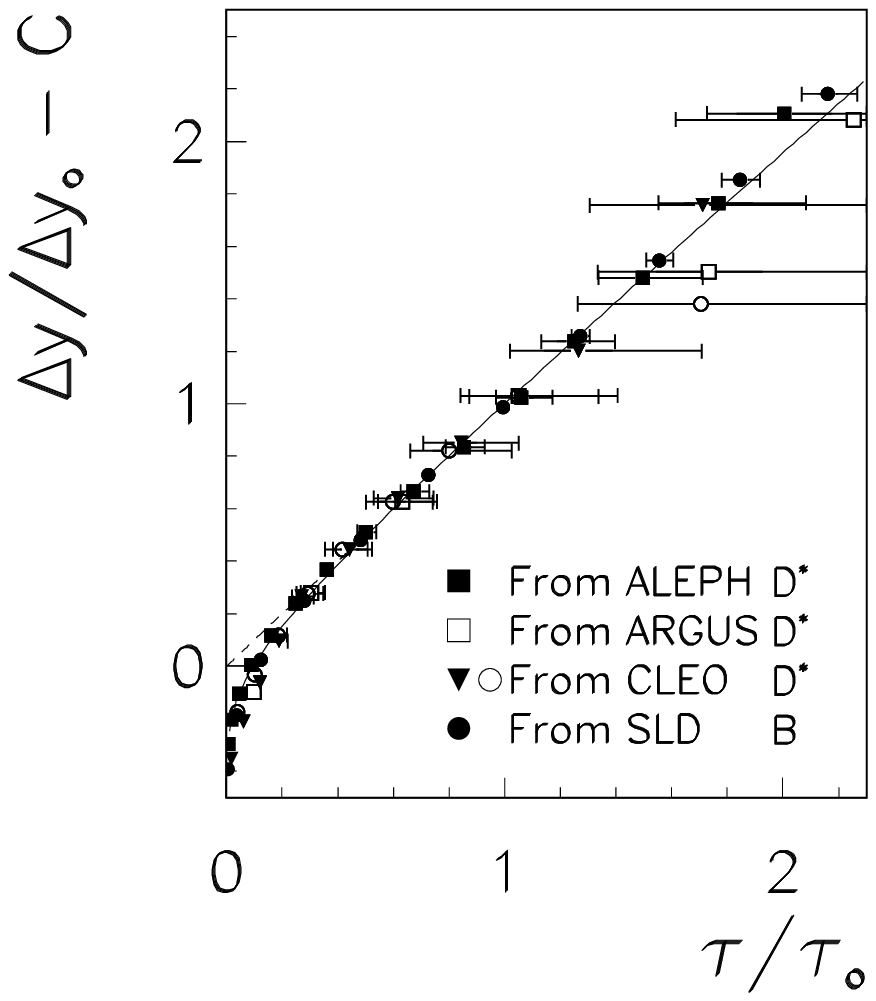,width=10.0cm}}
\put(7.1,-0.5)
{\epsfig{file=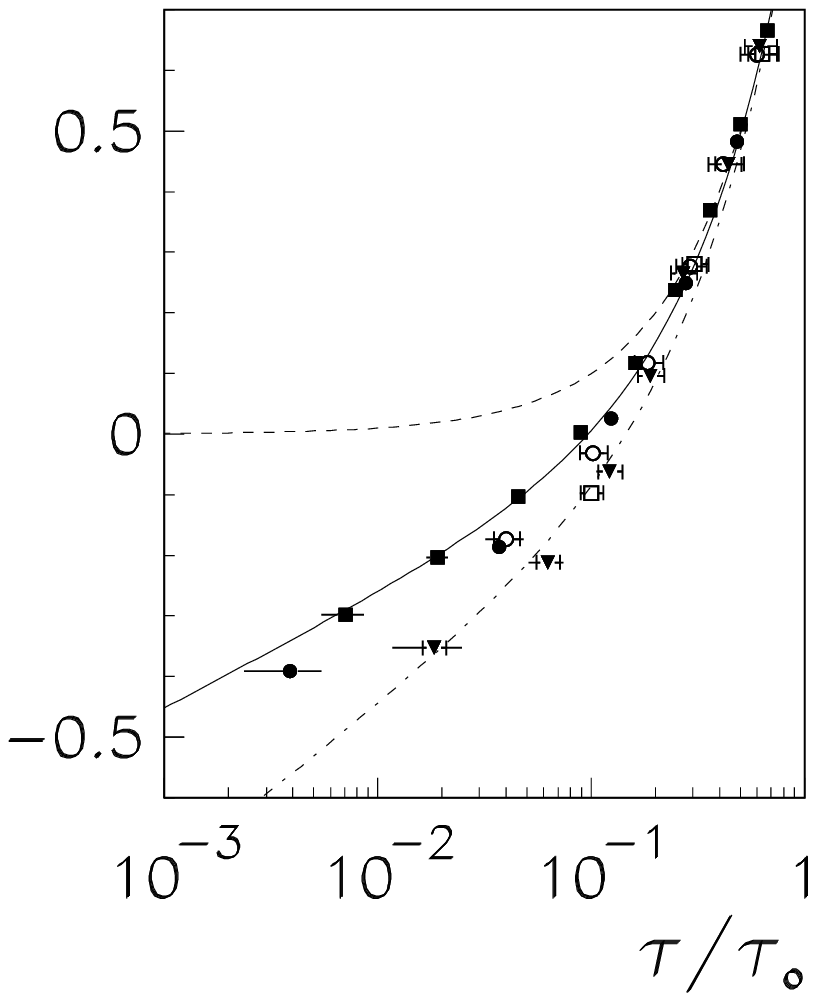,width=10.0cm}}
\end{picture}
\caption{The rapidity change $\Delta y$ of charm and bottom quarks,
measured by D$^*$ and B mesons from $e^+e^-$ collisions, 
is shown as a function of the quark proper time $\tau$.
The dashed curve denotes a linear dependence of $\Delta y$ on $\tau$
as expected for a constant force.
The full curve represents a logarithmic dependence of $\Delta y$
on $\tau$ for the short distance part, and a linear
dependence for the long distance part. 
The dash-dotted curve shows an increased strength of the logarithmic term.}
\label{fig:v} 
\end{figure}

In Fig.~\ref{fig:v}, the horizontal axis shows the proper time $\tau$
in units of $\tau_\circ$ from solving (\ref{eq:etau}).
The vertical axis gives the rapidity change $\Delta y$.
It is multiplied by $1/\Delta y_\circ$
to account for the observed scaling of the rapidity change spectra
reported in section~\ref{sec:rap}.
The values of $C$
are similar in the different data sets (Table~\ref{tab:fits}).
In Fig.~\ref{fig:v}a the long distance part is emphasized.
It shows the linear relation between $\Delta y$ and $\tau$
which confirms the method of determining $N_{norm}/N_{q,\circ}$.
At small distances, the rapidity loss is much larger per unit time.

Owing to the integration procedure, the points within a single experiment
are correlated.
Two errors are shown. The terminated errors give the range 
allowed by the uncertainty in $N_{norm}/N_{q,\circ}$ and the other errors
result from the errors of the original measurements.
A simulation of the procedure showed that the simulated points 
reproduce the generated $\Delta y(\tau)$ distribution except for 
a small overall shift in $C$ from the integration using finite bin widths, 
and for the first point of the integration.
This lowest point was omitted from the figure if the simulated 
displacement in $\tau/\tau_\circ$ exceeded $100\%$.

Figure~\ref{fig:v}b shows the rapidity change as a function of the proper
time on a logarithmic scale.
The dashed curve indicates a linear rapidity loss.
In the short distance region, the D$^*$ and B measurements at the Z resonance
are consistent with a dependence of $\Delta y$ on the logarithm of $\tau$.
The full curve attempts a parameterization as
\begin{equation}
\frac{\Delta y}{\Delta y_\circ} - C = 
a \; \ln{\left( \frac{\tau}{\tau_\circ} \right)}
+ b \; \left( \frac{\tau}{\tau_\circ} \right) + c 
\label{eq:para}
\end{equation}
with $a=0.08, b=0.9$, and $c=0.1$.
The D$^*$ measurements at the smaller beam energy $\sqrt{s}=10.6$~GeV
are also consistent with
a logarithmic dependence, albeit with a tendency towards a larger value of 
$a$ shown by the dash-dotted curve with $a=0.12$.
In principle, a larger value of $a$ is expected from the running of the
strong coupling constant.
Attempts to parameterize the dependence of the rapidity loss as
$\sim -a/(\tau/\tau_\circ) + b (\tau/\tau_\circ) + c$
did not describe the data.

Therefore, the heavy quark mesons carry information on the distance 
dependence of the strong force.
Plotting as in Fig.~\ref{fig:v} the quark kinematic change 
($\Delta y$) as a function of the distance scale ($\tau$)
we see that information on the behaviour of the strong interaction 
potential between the quark--antiquark pair can be extracted 
directly from the data.

\section{Fragmentation Distributions}

\noindent
In this section, the analysis is turned around and is used to
predict fragmentation functions for hadrons which contain the
primary quark and which have not decayed.
The expected distributions of the energy change are obtained by
folding the exponential proper time distribution (\ref{eq:tau})
with the rapidity change as a function of the proper time 
determined in Fig.~\ref{fig:v}.

\begin{figure}[htt]
\setlength{\unitlength}{1cm}
\begin{picture}(15.0,15.0)
\put(2.0,-0.5)
{\epsfig{file=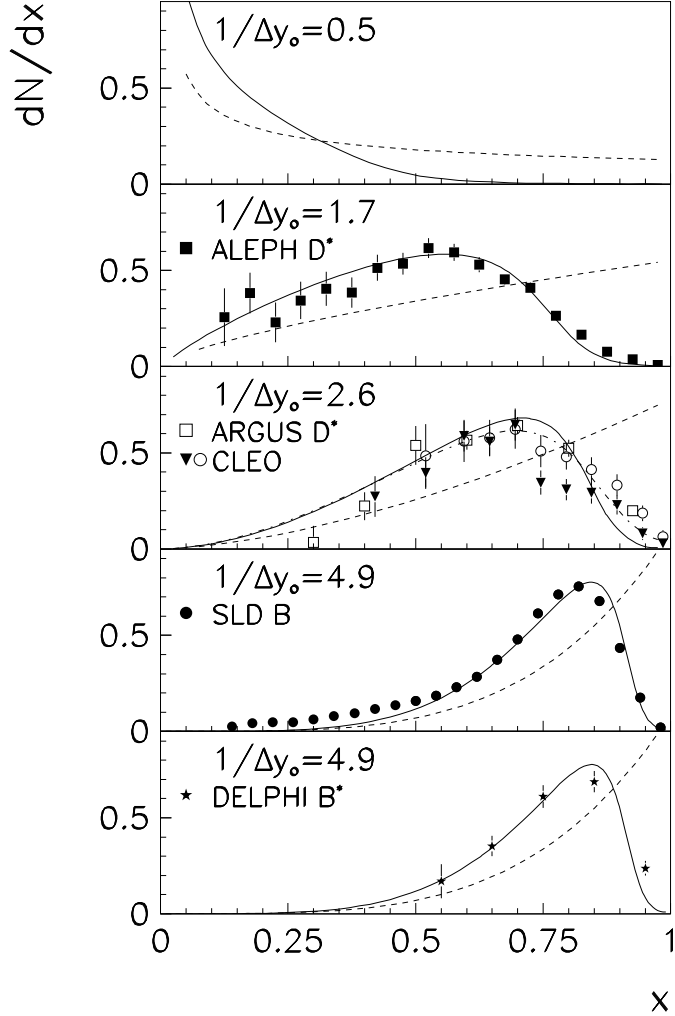,width=10.0cm}}
\end{picture}
\caption{Fragmentation functions are shown for 
different values of the scaling parameter $\Delta y_\circ$ 
and different situations:
the dashed curves denote results from a linear dependence of the
rapidity change on the time $\tau$.
The full curves give the energy loss of a logarithmic plus linear
dependence (\ref{eq:para}).
In the middle plot, the dash-dotted curve shows the effect of an 
increased strength in the logarithmic term.
The symbols represent the shapes of measured fragmentation 
distributions.}
\label{fig:x} 
\end{figure}

In Fig.~\ref{fig:x}, fragmentation functions
are shown for different values of the scaling parameter $\Delta y_\circ$.
The full curves give the situation where the quarks experience
a rapidity loss according to (\ref{eq:para}) together with $C=0.55$.
For the D$^*$ data at $\sqrt{s}=10.6$~GeV, the
dash-dotted curve shows the effect of the increased strength
in the logarithmic term.

It is instructive to also consider the situation where a linear relation
$\Delta y \propto \tau$, i.e. a constant force, 
is responsible for the fragmentation distributions
(dashed curves).
In this case, the rapidity distribution is exponentially falling
(eqs. (\ref{eq:acc}, \ref{eq:con}, \ref{eq:tau})),
leading to energy distributions of the power law form
\begin{equation}
\frac{dN}{dx} \sim x^{\left[ 1 / \Delta y_\circ - 1 \right]}
\sim x^{\left[ m / (\kappa \tau_\circ) - 1 \right]} \; .
\end{equation}
For accelerated masses $m=\kappa \tau_\circ$,
the $x$-distributions are expected to be flat.
For $m>\kappa \tau_\circ$ and $x\rightarrow 1$ 
the $x$-distributions are expected to increase 
and for $m<\kappa \tau_\circ$ to decrease.
As an example of the latter case, the
upper figure shows the prediction for strange mesons, 
assuming $1/\Delta y_\circ=0.5$ 
which corresponds to $m=0.25$~GeV at $\kappa \tau_\circ=0.5$~GeV.

As a consequence of choosing $x$ as the standard fragmentation
observable for the hadrons considered here, 
light and heavy flavour fragmentation processes seem to be very different.
Alternatively, using $\Delta y$ as the fragmentation observable
the spectra are expected to differ only by a scaling parameter.

\section{Concluding Remarks}

\noindent
Measurements of heavy quark production in $e^+e^-$ collisions 
have been analysed in terms of the quark rapidity change.
The D$^*$, B$^*$, and B meson spectra are found to follow a universal shape 
when expressed in units of a scaling parameter.

The shapes of the measurements are suggestive
of a hadronization mechanism which is probabilistic in nature.
Taking the view that the probability of hadron formation is constant in the rest 
frame of the travelling quark, the accelerated masses were determined.
In principle, the measurements could be much improved 
by dedicated analyses using the original event data.
A determination of the strange quark mass is desirable from 
measurements of kaons which are directly produced in 
$e^+e^-\rightarrow s\bar{s}$ processes.

Furthermore, the mesons carry information on the variations 
in the strength of the quark--antiquark interactions.
Directly from the data, information on the behaviour 
of the strong interaction potential 
between quark--antiquark pairs was extracted.
It remains for ways to be found 
to compare the measurements to QCD lattice calculations.
It will be of great interest to determine the potential between quarks
and di-quarks from measurements involving protons.
Comparisons of baryon spectra  
in central and peripheral collisions of 
heavy nuclei may be used in searches for the quark--gluon plasma where
the field strengths are expected to be different.

Within the perspective of colour charged objects, losing energy in a known
strong field for some time before they hadronize, fragmentation distributions 
of particles which contain the primary quark and which have not decayed
can be viewed as essentially depending on the mass accelerated 
in the field.

\section*{Acknowledgements}
For very constructive, fruitful discussions and comments I wish to thank 
G.~Barker, J.~Dainton, M.~Feindt, S.~Kappler, P.~Schleper, and T.~Walter.
I wish to thank Th.~M\"uller and the IEKP group of the University
Karlsruhe for their hospitality, and the Deutsche Forschungsgemeinschaft 
for the Heisenberg Fellowship.


\begin{thebibliography}{00}
\bibitem{cleo} 
CLEO Collab., D.~Bortoletto et al., {\em Phys. Rev.\/} {\bf D37} (1988) 1719
\bibitem{argus}  
ARGUS Collab., H.~Albrecht et al., {\em Z. Phys.\/} {\bf C52} (1991) 353
\bibitem{aleph}
ALEPH Collab., R.~Barate et al., {\em Eur. Phys. J.\/} {\bf C16} (2000) 597 
\bibitem{sld} 
SLD Collab., Kenji Abe et al., {\em Phys. Rev. Lett.\/} {\bf 84} (2000) 4300
\bibitem{delphi} 
DELPHI Collab., P.~Abreu et al., {\em Z. Phys.\/} {\bf \/} {\bf C68} (1995) 353
\bibitem{eichten} E.~Eichten et al., 
{\em Phys. Rev. Lett.\/} {\bf 34} (1975) 369, \\
{\em Phys. Rev.\/} {\bf D17} (1978) 3090 and {\bf D21} (1980) 203
\bibitem{bde} W.~Busza, T.~Dreyer, M.~Erdmann, {\em Z. Phys.\/} {\bf C48} (1990) 167
\bibitem{de}    T.~Dreyer, M.~Erdmann, {\em Phys. Lett.\/} {\bf B260} (1991) 232
\bibitem{pdg} 
D.E.~Groom et al., {\em Eur. Phys. J.\/} {\bf C15} (2000) 1 
\bibitem{delphi-bmass}
DELPHI Collab., P.~Abreu et al., {\em Phys. Lett.\/} {\bf B418} (1998) 430
\bibitem{brandenburg}
A.~Brandenburg et al., {\em Phys. Lett.\/} {\bf B468} (1999) 168 
\bibitem{aleph-bmass}
ALEPH Collab., R.~Barate et al., {\em Eur. Phys. J.\/} {\bf C18} (2000) 1
\end{thebibliography}
\end{document}